\documentclass[aps,pre,twocolumn,amsmath,amssymb]{revtex4} 
\usepackage{graphicx}  
\usepackage{dcolumn}   
\usepackage{bm}        

\newcommand{\eq}[1]{Eq.~(\ref{#1})}
\newcommand{\fig}[1]{Fig.~\ref{#1}}
\newcommand{\keff}{K_{\rm eff}}

\begin{document}
\title{A modified Potts model for the interaction of surface-attached polymer complexes}
\author{Marcel Hellmann$^{1,3}$, Youjin Deng$^2$, Matthias Weiss$^1$, and Dieter W. Heermann$^3$}
\affiliation{$^1$ Cellular Biophysics Group, German Cancer Research
	Center, Im Neuenheimer Feld 280, D-69120 Heidelberg, Germany}
\affiliation{$^2$ Physikalisches Institut, Philosophenweg 12, Universit\"at
 	Heidelberg, D-69120 Heidelberg, Germany}
\affiliation{$^3$ Institut f\"ur Theoretische Physik, Philosophenweg 19, Universit\"at
 	 Heidelberg, D-69120 Heidelberg, Germany}

\begin{abstract}
We present a simple yet generic model for the behavior of a system of many 
surface-attached flexible polymers with rigid side chains. Beyond its 
potential application in describing the dynamics of the extracellular matrix 
of mammalian cells, the model itself shows an interesting phase transition 
behavior since the underlying models (a two-dimensional Potts model and a 
XY-model) undergo different phase transitions.
\end{abstract}
\maketitle

\section{Introduction}
Surface-attached polymers are an ubiquitous phenomenon in living matter. On
the cellular level, prominent examples are the cytoskeletal actin 
network beneath the cell's plasma membrane \cite{alberts} and the extracellular 
matrix \cite{LLJ93}. While the semiflexible actin polymers interact at
multiple sites with the intracellular face of the cell's membrane, the
dominant component of the extracellular matrix, the hyaluronic acid (HA), 
rather is a flexible polymer, end-grafted to the extracellular face of the
membrane and extending to the cell's environment. To provide an efficient
barrier that is capable of protecting the cell, extracellular HA is modified
by rigid aggrecan combs that are linked to the flexible HA backbone. As a
consequence, the cell is protected by a 'jungle' of HA-aggrecan complexes.

We have previously shown that attaching a single rigid side chain to a
flexible, end-grafted polymer (resembling the HA-aggrecan complex) can
lead to a considerable stiffening of the backbone \cite{HWH07}, hinting 
at a mechanism by which cells can tune the rigidity of their protective matrix.
In this study, we had used a dynamic Monte Carlo algorithm on a regular cubic
lattice to study the steady-state properties of a flexible, end-grafted
backbone polymer with a rigid side chain. Extending this approach to a system
of many interacting backbones, however, is computationally challenging due to 
the massive increase in the system's autocorrelation time.

An alternative approach to a brute-force Monte Carlo sampling is a further
coarse graining of the system that reduces the problem to its basic
observables, namely the height of the backbones and the orientation of the
rigid side chains. Here, a combination of the Potts model (describing the
height levels) and the two-dimensional XY-model (describing the orientation of
the side chains) appears as a promising candidate. Using this approach, however, 
one has to deal with an intricate problem of statistical physics -- the
prediction of the system's phase behavior and phase transition. The Potts
model (with $q<4$ height levels) undergoes a second order phase 
transition \cite{Baxter1973} while the two-dimensional XY-model shows a 
Kosterlitz-Thouless transition \cite{Kosterlitz1973,Kosterlitz1974}. It is far 
from evident which phase transition will be encountered when coupling these 
two systems. Aiming at modeling the behavior of the extracellular matrix in 
terms of such a combined model thus requires to elucidate the model's phase 
behavior in the first place.

Inspired by the potential value of the combined Potts-XY-model in describing
features of the extracellular matrix, we have investigated the phase
transition of the model. In particular, we have studied the phase transition 
in a model that couples the $q=3$ Potts model with the XY-model by means of
extensive Monte Carlo simulations. We find that the phase transition is 
dominated by the Potts model, i.e. a second order phase transition is
observed. The ordered state at low temperatures shows domains of parallel
aligned XY-spins with the Potts levels collectively exploring all states in a
stochastic fashion. In contrast, for high temperatures the XY-spins are
disordered with an average Potts level $\langle q\rangle=1$ that only varies
to a minor extent. Relating these findings to the original biological problem, 
we speculate that changing the interaction strength between individual
HA-aggrecan complexes, i.e. driving the system through the phase transition,
may be one way for the cell to build up a stochastically fluctuating
protective barrier. 

\section{Model}
\begin{figure} 
\begin{center}
  \includegraphics[width=7cm]{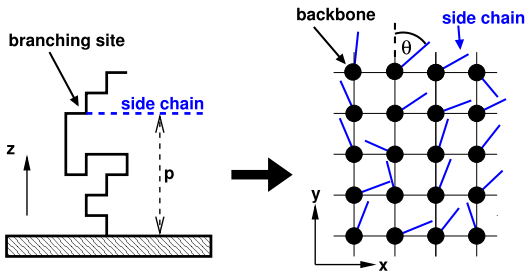}
  \caption{For describing as system of end-grafted, self-avoiding polymer 
    backbones with rigid side chains (left), only the essential degrees of
    freedom, i.e. the height $p$ of the side chain above the substrate
    and its orientation angle $\theta$, are retained. A system of mutually
    interacting complexes is obtained by using a two-dimensional lattice of 
    such rotors (right) where the dynamics is determined by \eq{eq:Hybrid_model}.    
} 
\label{fig:fig01}
\end{center}
\end{figure}

As described previously \cite{HWH07}, the basic unit of the
extracellular matrix may be reduced to a single flexible, end-grafted
backbone with a rigid side chain (\fig{fig:fig01}). While a simulation of 
this unit yields valuable insights into the stiffening of the individual 
backbone due to the attached rigid side chain, many of the details
may not be necessary to understand the generic behavior of 
the multi-complex system. 
The basic quantities describing the interaction of many individual units 
are the height $p$ of each side chain above the substrate and its orientation 
angle $\theta$ with respect to the $y$-axis.
Thus, the individual backbone side-chain complexes are interpreted as
mutually interacting rotors aligned on a regular two-dimensional lattice. 

We assume here that the sites at which the backbones are attached to the substrate
are sufficiently far aparat from each other so that any interaction between
the complexes is mediated by the side chains only. This also means that
side chains only interact when they have the same height $p$. This aspect can 
be modeled by a Potts-type Hamiltonian \cite{Potts1952,Wu1982}:
\begin{equation} \label{eq:Potts_model}
\mathcal{H}_{P} = - K_P\sum \limits_{\langle i,j \rangle}\delta_{p_i p_j}\,\,; \quad p_{i} \in \{0,1,\dots,q-1\}\,\,.
\end{equation}  
Here, $p_i$ denotes the height of side chain $i$ above the substrate, $q$ is the
total number of allowed heights, and $K_P>0$ is the coupling constant (in units
of $\beta = k_B T$). 

The interaction between side chains with the same $p$ will be considered as a 
nearest-neighbor ferromagnetic interaction leading to a prefered parallel
alignment of the side chains. This kind of interaction seems plausible as a 
first approach since it mimics the repulsive forces between side chains that 
may be due to steric or short-ranged electrostatic potentials. Consequently,
we use for this part of the interaction the two-dimensional XY-model Hamiltonian:
\begin{equation} \label{eq:XY_model}
	\mathcal{H}_{XY} = - K_{XY}\sum\limits_{\langle i,j \rangle} 
\mathbf S_i \mathbf S_j = -K_{XY} \sum \limits_{\langle i,j \rangle} S_i S_j \cos \theta_{ij}\,\,,
\end{equation}
with the coupling constant $K_{XY}>0$ and $\langle i,j \rangle$ denoting a
summation over nearest neighbors. For simplicity, the length of the rotors
(=side chains) $S_i = \vert \mathbf S_i \vert$ is set to unity while the 
angle $\theta_{ij}$ denotes the relative orientation of the side chains $i$
and $j$. 

Combining the above models \eq{eq:Potts_model} and \eq{eq:XY_model} yields the 
hybrid model that is used to describe the surface-attached polymer complexes
of the extracellular matrix:
\begin{equation} \label{eq:Hybrid_model}
\mathcal{H} = -K \sum \limits_{\langle i,j \rangle} 
\delta_{p_i p_j} \, S_i S_j \cos \theta_{ij}\,\,;\quad  p_{i} \in \{0,\dots,q-1\}\,\,.
\end{equation}
The two-dimensional Potts model is known to undergo a second order phase transition
for $q<4$ \cite{Baxter1973} while the XY-model shows a Kosterlitz-Thouless transition 
\cite{Kosterlitz1973,Kosterlitz1974}. For $q=3$ our model \eq{eq:Hybrid_model}
can be interpreted as a 3-state Potts model with a randomly varying coupling
strength $K_{ij} = \cos \theta_{ij}$ between neighbouring Potts spins $i$ and
$j$. As this local coupling may become very small and even zero, the nature of
the phase transition of the hybrid model may be influenced by \emph{vacancies}
that are known to drive the transition of the ($q<4$)-state Potts model to a 
first order behavior \cite{Nienhuis1979}. Given this variety of phase behaviors,
it appears difficult to predict the system's behavior \emph{a priori}.
\begin{figure} 
\begin{center}
  \includegraphics[width=7cm]{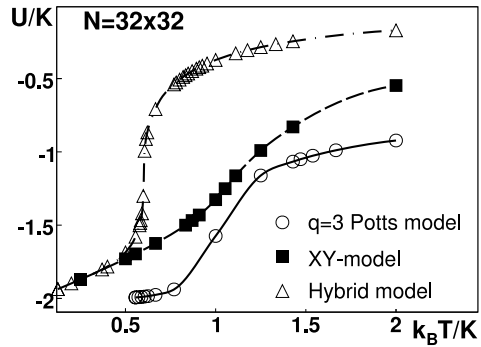}
  \caption{The inner energy $U$ per rotor shows a sigmoidal increase with temperature
    $T$ for the pure XY-model [\eq{eq:XY_model}], the pure $3$-state Potts
    model [\eq{eq:Potts_model}], and the hybrid model [\eq{eq:Hybrid_model}].
    The transition temperature $T_c$ for the latter appears to be shifted
    towards lower values while the slope in the transition region is strongly
    increased. Full lines are guides to the eye.}
\label{fig:fig02}
\end{center}
\end{figure}

\section{Simulation Method}
For simulations of the hybrid model \eq{eq:Hybrid_model} we relied on the Monte Carlo 
(MC) method (see, e.g., \cite{Heermann2002} for an introduction). Far away from the
critical temperature, we employed the Metropolis MC algorithm
\cite{Metropolis1953}. In each step, a rotation of each rotor (side chain) 
by some angle $\Delta \theta_i$ and a change of its Potts level (height) 
were proposed and this combined move was accepted or rejected for each rotor 
according to the Metropolis criterion.

Close to the critical temperature, the standard Metropolis MC suffers from 
critical slowing down, i.e. it becomes difficult to generate enough
statistically independent configurations for a reliable statistics. To
overcome this problem, we used a cluster MC approach \cite{Swendson1987}, i.e.
a modification of Wolff's cluster algorithm \cite{Wolff1989}. Here, 
one deals with two degrees of freedom, one discrete (Potts level $p$) and one 
continuous (orientation angle $\theta$). Thus, two different kinds of clusters 
are grown and flipped in each step -- one with respect to $p$ (embedded in an 
ensemble of fixed $\theta_i$'s) and another one with respect to $\theta$
(embedded in an ensemble of fixed $p_i$'s).

The treatment of the Potts degree of freedom has to be carried out carefully 
as ferromagnetic and anti-ferromagnetic bonds can occur between neighboring 
rotors. In the beginning, a pair of values for $p$ is chosen randomly, i.e. 
(0,1), (0,2), or (1,2), respectively. Rotors in the remaining Potts state 
are kept unchanged, i.e.,the identity operator is applied, and thus
do not contribute to the ongoing cluster growing step. The effective 
(local) coupling between rotors $i$ and $j$ is $\keff = K \cdot S_i S_j \cos \theta_{i,j}$.
For $\keff>0$ a ferromagnetic bond is established with probability
$P = 1 - \exp \left( -\keff \delta_{p_i,p_j} \right)$
between rotors on the same Potts level. If $\keff<0$, an anti-ferromagnetic 
bond is formed with probability$P = 1 - \exp \left( \keff (1-\delta_{p_i,p_j})  \right)$
between rotors on different Potts levels. When the cluster is grown, all Potts
levels of the involved rotors are switched according to the initially
chosen pair of $p$-values (e.g. 0 $\leftrightarrow$ 1).

Following Wolff's embedding trick for growing and flipping clusters in the XY-model
\cite{Wolff1989} the rotors are projected onto the $x$-axis. This
results in an Ising model of the $x$-components $S^x_i$  with random
couplings. These 'effective' spins are connected by a bond with probability
$P=1-\exp\left(\delta_{p_i,p_j}\cdot\min [0,2K S^x_i S^x_j ] \right).$
Please note that the Potts states of the rotors are accounted for by
$\delta_{p_i,p_j}$ meaning that bonds are only formed between rotors on the
same Potts level. Finally, when the cluster is grown, all $x$-components of the
constituing rotors are inverted.
\section{Results}
To study the thermodynamic properties of a system of interacting rotors
modelled by the Hamiltonian \eq{eq:Hybrid_model} with $q=3$ we conducted
extensive MC simulations using the method described in the preceeding
section. The linear system size $L$ (i.e. using $N=L^2$ rotors) was varied in
the range $32\le L\le 1024$ in order to apply a finite-size scaling
analysis. We have concentrated here in particular on the inner (potential) 
energy $U$ per rotor and the transition temperature $T_c$.
\begin{figure}
\begin{center}
  \includegraphics[width=7cm]{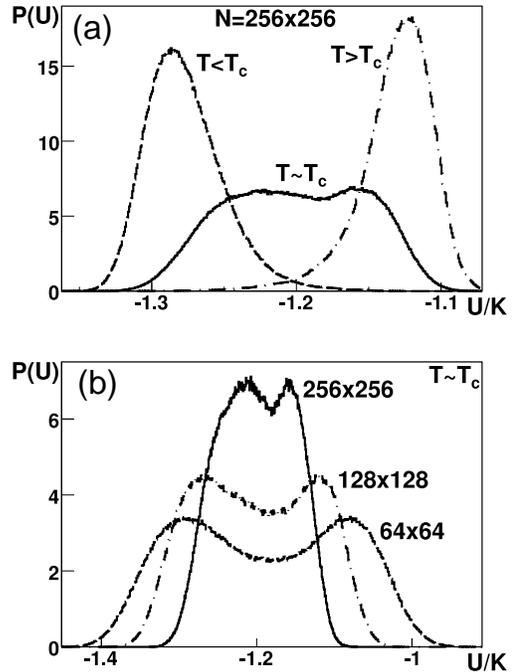}
  \caption{(a) The inner energy $U$ below and above the critical temperature
    $T_c$ shows only a single maximum for a system with $N=256$, while for
    $T\approx T_c$ a slight double-peak structure is observed. (b) Indeed, this
    double-peak structure at $T\approx T_c$ is even better visible for small 
    systems and seems to subside for larger system sizes.} 
\label{fig:fig03}
\end{center}
\end{figure}

Metropolis MC simulations of small systems revealed significant differences in
the behavior between the hybrid model and the two constituting models (Potts
and XY). \fig{fig:fig02} shows the variation of $U$ with temperature for all
three models. Obviously, the transition point $T_c$ of the hybrid model is
much lower than those of the constituting models. Furthermore, the transition
region is much steeper, suggesting a first-order transition. At this point,
however, it cannot be determined if the slope really diverges at $T_c$ which 
would be a criterion for a discontinuous phase transition. This behavior would 
be qualitatively different from that shown by the constituting models, yet
would agree with the random Potts model \cite{Nienhuis1979}. 
Please note that for $T<T_c$ the energy $U$ of the hybrid model resembles
essentially that of the corresponding pure XY-model. Thus, the Potts degree 
of freedom seems to be 'frozen' in this regime while all excitations occur
with respect to the XY-degree of freedom.
\begin{figure} 
\begin{center}
  \includegraphics[width=7cm]{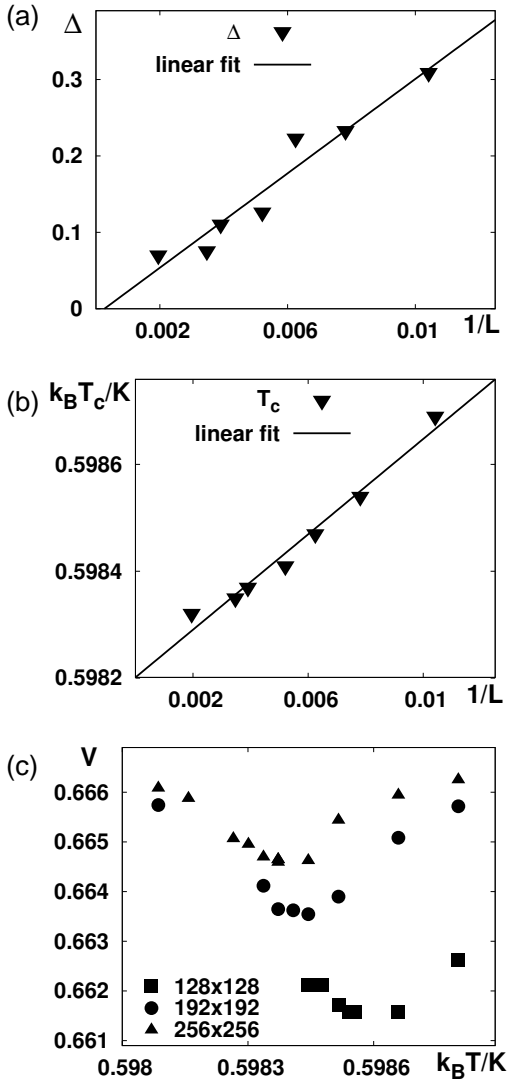}
  \caption{(a) The depth $\Delta=\ln[P(U)_{max}/P(U)_{min}]$ of the minimum 
   between the double peak in $P(U)$ appears to vanish with system size $L$,
   thus indicating a second-order phase transition. (b) In agreement with this, 
   the phase transition temperature shows a scaling $T_c\sim 1/L$. (c) The
   tendency of the 4$^{th}$-order cumulant towards a value $V=2/3$ is also
   in support of a second-order phase transition.}
\label{fig:fig04}
\end{center}
\end{figure}

For a more thorough investigation, we used the cluster algorithm described in
the preceeding section. In contrast to the standard Metropolis MC method it
allowed for extensive simulations of large systems close to $T_c$ due
to a reduced dynamic exponent. We focussed on the inner energy $U$ as we found
no proper order parameter to follow the phase behavior of the hybrid model. As
can be seen from \fig{fig:fig03}a, the distribution $P(U)$ shows pronounced 
single peaks for $T\ll T_c$ and $T\gg T_c$, while for $T\approx T_c$ a double-peak
seems to emerge. In \fig{fig:fig03}b, the distribution $P(U)$ is shown for
three different system sizes at $T \approx T_c$. The distributions were
obtained by \emph{reweighting} \cite{Ferrenberg} of distributions close to
$T_c$. Our numerical data indeed highlight a double-peak structure in $P(U)$, 
indicating the coexistence of two phases and thus a first-order phase transition.
To extrapolate if this finding persists in the thermodynamic limit, we
first considered the logarithm of the relative gap depth $\Delta=\ln [P(U)_{max}/P(U)_{min}]$ 
at $T \approx T_c$ for growing system sizes~$L$. For a first-order transition, 
this quantity is supposed to grow with $L$. In the hybrid model studied here, 
however,  $\Delta\to0$ for $L \to \infty$ (\fig{fig:fig04}a). Thus, the energy 
distribution assumes a Gauss-like shape in the thermodynamic limit, consistent 
with a second-order phase transition with only a single phase existing in the 
system at $T_c$. Considering the finite-size scaling of the transition
temperature $T_c$ with the inverse system size $1/L$ supports this notion 
since we find $T_c \propto 1/L$ as anticipated for a second-order transition
(\fig{fig:fig04}b). 

Further evidence for a second order phase transition comes from the 4$^{th}$
order cumulant of the energy distribution \cite{Binder1981}
\begin{equation}
	V = 1 - \frac{\langle U^4  \rangle}{3 \cdot \langle U^2 \rangle^2}\,\,.
\end{equation}
For a Gaussian distribution, this quantity takes on the value $V=\frac{2}{3}$
while distributions with a double-peak structure are characterized by 
$V\neq\frac{2}{3}$. While a small dip in $V(T)$ is observed for small system 
sizes (\fig{fig:fig04}c), it subsides for large $L$ and only very small
deviations from $V=2/3$ are seen. In the thermodynamic limit, one thus 
expects $V=\frac{2}{3}$, i.e. a Gauss(-like) distribution for all temperatures.

\begin{figure} 
\begin{center}
  \includegraphics[width=7cm]{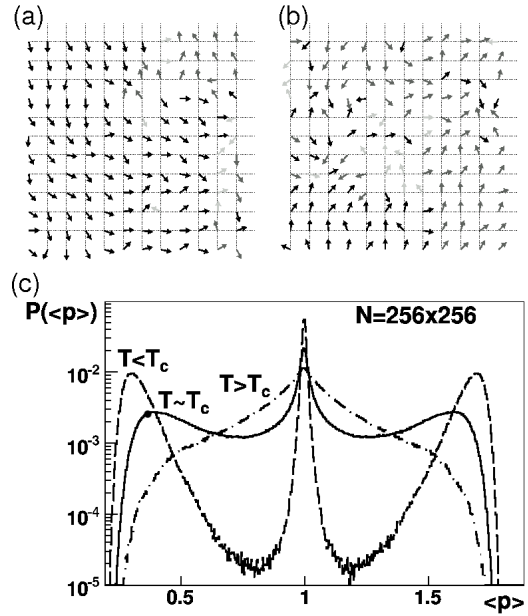}
  \caption{Snapshot of the system at (a) $T<T_c$  and (b) $T>T_c$. Rotor
  orientations are indicated by arrows, Potts levels $p=1,2,3$ are labeled 
  in black, dark grey and light grey, respectively. (c) Probability
  distribution of the mean Potts level in the system $P(\langle p\rangle)$.
  In the ordered state at $T<T_c$ (dashed), all Potts states $p=1,2,3$ are 
  populated with similar probabilities while for $T\approx T_c$ (dash-dotted) 
  and $T>T_c$ (full) the state $p=1$ is clearly favored.}
\label{fig:fig05}
\end{center}
\end{figure}

The phase behavior of the system above and below the critical temperature is
shown in \fig{fig:fig05}. While for $T<T_c$ the rotors (=rigid side chains) are 
ordered (\fig{fig:fig05}a), they assume random orientations for $T>T_c$. The
mean Potts level (=the average height of the rigid side chains) in both cases
is $\langle q\rangle=1$ (\fig{fig:fig05}c), yet for $T<T_c$ collective 
fluctuations of the rotors to the three degenerate Potts levels are observed. 
Given the small transition probabilities for these height fluctuations for 
fixed temperatures (which we circumvented via the MC cluster algorithm) the 
dwell time at an individual Potts levels may be very large.
\section{Conclusions}
In this work, we have proposed a model for a system of surface-attached polymers
with rigid side chains that is found, e.g., in the extracellular
matrix of mammalian cells. To reduce the computational effort when simulating 
large systems of end-grafted flexible polymers with rigid side chains, a
coarse grained approach has been chosen combining two well-known models of 
statistical physics, namely the Potts and the XY-model. The thermodynamics of 
the 'hybrid' model, given by the Hamiltonian \eq{eq:Hybrid_model}, has been
studied in Monte Carlo simulations and a second-order phase transition was found.

Given our results, it seems reasonable to call the hybrid model rather a
'modified Potts model' than a 'modified XY-model' as the phase behavior of the 
system is dominated by the 3-state Potts model: At low temperatures, the
rotors essentially condensate into a single Potts state. This corresponds to 
the (three-fold degenerate) ground state of the Potts model showing
long-range order with respect to $p$. Excitations in this regime arise due to 
perturbations in the ordered state of the XY-model. At high temperatures, 
the influence of the XY degree of freedom becomes manifest as a randomized
coupling of the Potts spins which includes the existence of vacancies
that may drive the transition of the 3-state Potts model from second to first 
order \cite{Nienhuis1979}. Indeed, at first glance, the transition of the
hybrid model appears to change character (cf. \fig{fig:fig02} and \fig{fig:fig03}).
Only more thorough investigations revealed the transition to be still 
continuous, i.e. to remain second order. Thus, the vacancies evoked by the
random coupling drive the transition towards first order, but do not suffice
to fully change its character.

Coming back to the original problem that inspired the model, i.e. 
the dynamics of the extracellular matrix, the observed ordering of the system 
for $T<T_c$ may actually be used by a cell to change the thickness of its
protective layer. A cell may use the transition to the regime $T<T_c$ to
freeze the extracellular matrix in a desired homogenous ordered state/height
that is not necessarily 'thicker' than in the disordered regime. Thus,
speaking in terms of the original problem, the cell may tune the thickness of 
its extracellular matrix by tuning the interactions of the (polyelectrolytic) 
aggrecan side chains, thereby driving the system through the phase transition.
It will be interesting to experimentally manipulate the aggrecan interaction 
by changing ion conditions for the extracellular matrix and thereby to induce 
the above described phase transition.

\begin{acknowledgments}
This work was supported by the Institute for Modeling and Simulation in
the Biosciences (BIOMS) in Heidelberg. MH acknowledges financial support by
the Helmholtz alliance {\em SBCancer}. YD thanks the Alexander von Humboldt 
Foundation.
\end{acknowledgments}

\end{document}